# Medical Slice Transformer: Improved Diagnosis and Explainability on 3D Medical Images with DINOv2

Gustav Müller-Franzes[1], Firas Khader[1], Robert Siepmann[1], Tianyu Han[1], Jakob Nikolas Kather[2,3,4], Sven Nebelung[1*], Daniel Truhn[1*]

[1]Department of Diagnostic and Interventional Radiology, University Hospital Aachen, Aachen, Germany
[2]Else Kroener Fresenius Center for Digital Health, Technical University Dresden, Dresden, Germany
[3]Department of Medicine I, University Hospital Dresden, Dresden, Germany
[4]National Center for Tumor Diseases (NCT), University Hospital Heidelberg, Heidelberg, Germany
[*]Contributed equally

## Abstract

**Background**: MRI and CT are essential clinical cross-sectional imaging techniques for diagnosing complex conditions. However, large 3D datasets with annotations for deep learning are scarce. While methods like DINOv2 are encouraging for 2D image analysis, these methods have not been applied to 3D medical images. Furthermore, deep learning models often lack explainability due to their "black-box" nature. This study aims to extend 2D self-supervised models, specifically DINOv2, to 3D medical imaging while evaluating their potential for explainable outcomes.

**Method**: We introduce the Medical Slice Transformer (MST) framework to adapt 2D self-supervised models for 3D medical image analysis. MST combines a Transformer architecture with a 2D feature extractor, i.e., DINOv2. We evaluate its diagnostic performance against a 3D convolutional neural network (3D ResNet) across three clinical datasets: breast MRI (651 patients), chest CT (722 patients), and knee MRI (1199 patients). Both methods were tested for diagnosing breast cancer, predicting lung nodule dignity, and detecting meniscus tears. Diagnostic performance was assessed by calculating the Area Under the Receiver Operating Characteristic Curve (AUC). Explainability was evaluated through a radiologist's qualitative comparison of saliency maps based on slice and lesion correctness. P-values were calculated using Delong's test.

**Results**: MST achieved higher AUC values compared to ResNet across all three datasets: breast (0.94±0.01 vs. 0.91±0.02, P=0.02), chest (0.95±0.01 vs. 0.92±0.02, P=0.13), and knee (0.85±0.04 vs. 0.69±0.05, P=0.001). Saliency maps were consistently more precise and anatomically correct for MST than for ResNet.

**Conclusion**: Self-supervised 2D models like DINOv2 can be effectively adapted for 3D medical imaging using MST, offering enhanced diagnostic accuracy and explainability compared to convolutional neural networks.

# Main

Deep learning (DL) has demonstrated significant potential in medical imaging for diagnosis and analysis (1,2). However, its integration into clinical practice is hindered by the need for large, annotated datasets and limited model interpretability. Generating medical annotations is time-consuming and requires expert input (3,4). Furthermore, the "black-box" nature of DL models raises concerns about trustworthiness in clinical settings (5–7).

To enhance model explainability, techniques like saliency maps and Gradient-weighted Class Activation Mapping (Grad-CAM) (8) have been used to visualize network attention and provide insights into model decisions. However, these methods often lack accuracy, especially in 3D medical imaging (9,10). Transformers offer an alternative through their inherent attention mechanisms (11,12). Unlike convolutional models requiring post-hoc visualization, Transformers provide attention matrices highlighting the most relevant input features. Studies have shown that attention-based saliency maps improve interpretability in 2D medical imaging (13–15).

To address the scarcity of large annotated datasets, self-supervised learning combines labeled and unlabeled data to enhance model performance (16). Foundation models like DINOv2 (17) use self-supervised learning to extract features from vast amounts of unlabeled data. Despite being trained on natural images, these models have shown promise in transferring features to medical imaging tasks (18–24).

For instance, Huix et al. (18) found that DINOv2 outperformed other foundation models and even in-domain self-supervised ResNet152 in tasks like fundoscopy, mammography, dermoscopy, and chest radiography. Similarly, Truong et al. (23) and Nielsen et al. (24) demonstrated DINOv2's effectiveness on microscopic and endoscopic images. Huang et al. (19) reported that DINOv2 generally outperformed VGG and ResNet50 models pre-trained on ImageNet for chest radiographic, funduscopic, and dermoscopic images but not for brain MRI slices.

Despite these successes in 2D medical image analysis, the application of DINOv2 to 3D medical imaging remains unexplored. The present study aims to bridge this gap by introducing the Medical Slice Transformer (MST), a novel approach that adapts 2D self-supervised models like DINOv2 for 3D medical image analysis. We hypothesized that the MST framework would outperform standard 3D convolutional neural networks (such as ResNet) in terms of diagnostic accuracy across a set of clinical cross-sectional CT and MRI datasets. Furthermore, we hypothesized that MST would provide better localization information for identifying specific image findings, such as breast lesions, lung nodules, and meniscus tears, compared to ResNet.

The primary contributions of this study are as follows:

- Development of MST: We introduce MST, a technique designed to use self-supervised models like DINOv2 for 3D medical image analysis (**Figure 1)**.
- Enhanced Explainability: By leveraging the Transformer's attention mechanism, we demonstrate how model explainability can be improved, providing more insight into the decision-making process of DL models in 3D medical imaging contexts.

# Results

## Patient Characteristics

Three publicly available datasets were included: (i) breast MRI "Duke-Breast-Cancer-MRI" (DUKE) (25), (ii) chest CT "Lung Image Database Consortium and Image Database Resource Initiative" (LIDC-IDRI) (26), and (iii) knee MRI "MRNet" (27). Inclusion criteria were met by 651, 722, and 1199 patients for the breast MRI, chest CT, and knee MRI datasets, respectively (**Figure 1c**). The DUKE dataset comprised contrast-enhanced breast MR subtraction images from cancer patients and was used to evaluate the models' performance in breast cancer detection. The LIDC-IDRI dataset included chest CT scans with lung nodule annotations by four radiologists, allowing assessment of the models' ability to detect and analyze the dignity of lung nodules. The MRNet dataset contained knee MRI studies with meniscal tears, enabling evaluation of the models' effectiveness in diagnosing meniscal tears.

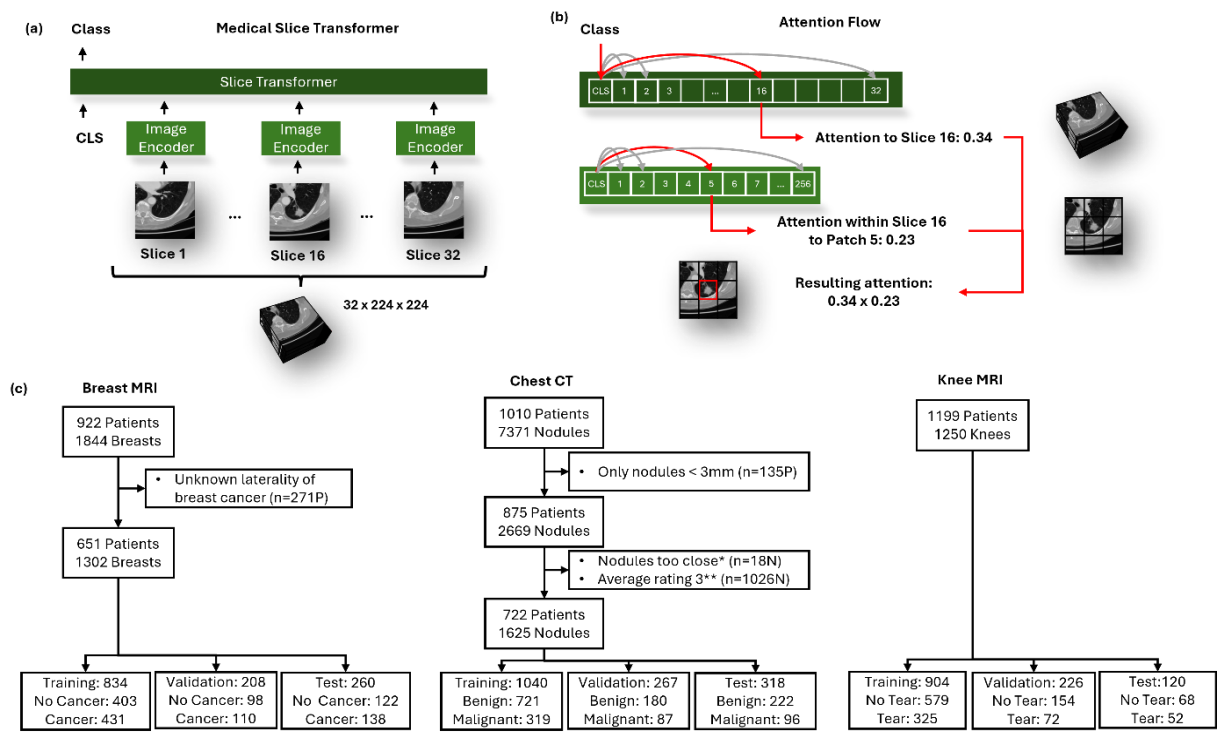

**Figure 1: Overview of the Model Architecture, Attention Flow, and Study Workflow**. (a) The Medical Slice Transformer framework processes individual MRI or CT slices using 2D image encoders, such as DINOv2, and then passes the encoded outputs through the Slice Transformer for downstream classification tasks. (b) Visualization of attention mechanisms showing how the Slice Transformer assigns attention to specific slices and how within-slice attention is further refined to specific patches, resulting in a combined attention map highlighting regions of interest in the input volume. (c) Study flow diagram of the Breast MRI dataset (Duke-Breast-Cancer-MRI), Chest CT dataset (Lung Image Database Consortium and Image Database Resource Initiative), and Knee MRI dataset (MRNet). Abbreviations: *Proximity of the nodules made them unresolvable by the pylidc library. **Radiologists rated the likelihood of malignancy on a 5-point scale ranging from "1: highly unlikely" to "5: highly suspicious"; thus, an average rating of "3" indicated unclear dignity. Abbreviations: CLS = classification token, P = Patient, N = Nodules.

## MST Outperforms Standard CNN Architectures

We compared the performance of a 3D ResNet50 (28) with the proposed MST architecture using DINOv2 as the image encoder by calculating the Area Under the Receiver Operating Characteristic Curve (AUC) for each dataset (**Figure 2**). MST exhibited higher AUC values compared to the 3D ResNet on all datasets: breast MRI: 0.94 ± 0.01 vs. 0.91 ± 0.02 (P = 0.02), chest CT: 0.95 ± 0.01 vs. 0.92 ± 0.02 (P = 0.13), and knee MRI: 0.85 ± 0.04 vs. 0.69 ± 0.05 (P = 0.001).

When systematically varying the reference model's complexity, we found that ResNet18 or ResNet101 did not improve performance over ResNet50 (**Supplementary Table S1**). ResNet18, ResNet50, and ResNet101 have 33 million, 46 million, and 117 million parameters, respectively. In contrast, the MST model has 23 million parameters, with 1 million parameters belonging to the Transformer part and 22 million to the DINOv2 part.

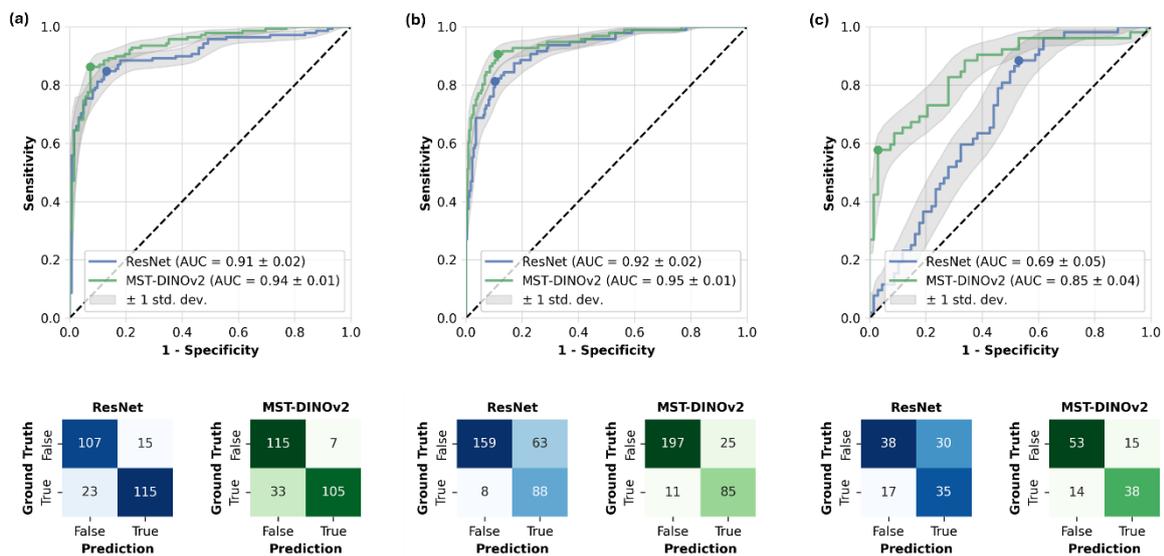

**Figure 2: Classification Performance as a Function of Model Architecture and Dataset.** Shown are the Area Under the Receiver Operating Characteristic Curve (AUC) values and confusion matrices for the Transformer-based self-supervised model (MST-DinoV2) and the standard convolutional neural network (ResNet) in the diagnosis of breast cancer in MRI (DUKE dataset, a), dignity assessment of lung nodules in CT (LIDC-IDRI dataset, b), and presence of meniscal tears (MRNet dataset, c).

## MST Provides Better Model Explainability

A major challenge with deep learning models in medical imaging is their lack of transparency, often called the "black-box" problem. For reference, we included the traditional Gradient-weighted Class Activation Mapping (Grad-CAM) to visualize the attention of the ResNet model.

To complement the quantitative analysis, we evaluated the saliency maps provided by both methods. A radiologist (R.S.) with five years of clinical experience performed a blinded evaluation of 50 saliency maps from each dataset, rating the models based on two key criteria (**Figure 3**):

    (i)      Slice correctness – does the saliency map highlight the slice(s) containing the lesion?
    (ii)     Lesion correctness – does the saliency map accurately point to the core of the lesion?

Irrespective of the clinical dataset, the saliency maps generated by the 3D ResNet were often distributed across multiple slices that did not contain lesions. They provided minimal information about the precise location of the different lesions (**Figures 4b, 5b, 6b**). The radiologist rated 37 of the provided 150 saliency maps as correctly highlighting the slice containing the lesion, and none (in any dataset) as accurately pointing to the core of the lesion **(Table 1)**.

In contrast, the slice-wise attention maps produced by the MST framework effectively highlighted relevant slices (**Fig. 4c, 5c, 6c**) and, when combined with the within-slice attention, pointed out the core of the lesions more reliably (**Fig. 4d, 5d, 6d**). The radiologist rated 136 of the provided 150 saliency maps as correctly highlighting the slice(s) containing the lesion, and 57 of 150 as accurately pointing to the core of the lesion.

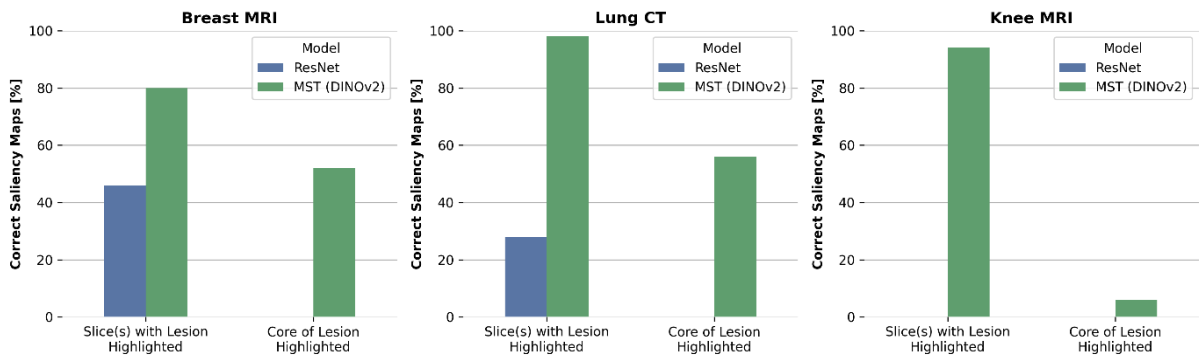

**Figure 3**: **Quantification of Saliency Map-Lesion-Correspondence as a Function of Imaging Dataset and Model Architecture.** Percentages of (blinded) radiologist evaluation in terms of slice correctness ("Does the saliency map highlight the slice(s) containing the lesion?" - yes/no) and lesion correctness ("Does the saliency map accurately point to the core of the lesion?" - yes/no).

**Table 1: Performance of 3D ResNet- and MST-Derived Saliency Maps in Highlighting Slices with Lesions and Cores of Lesions as a Function of Dataset.** "Slice Correctness" refers to the model correctly highlighting the slice(s) containing the lesion. "Lesion Correctness" refers to the model accurately pointing to the core of the lesion.

| Dataset | Method | Slice Correctness | Lesion Correctness |
|---|---|---|---|
| Breast | 3D ResNet | 23/50 | 0/50 |
| Lung | 3D ResNet | 14/50 | 0/50 |
| Knee | 3D ResNet | 0/50 | 0/50 |
| Breast | MST | 40/50 | 26/50 |
| Lung | MST | 49/50 | 28/50 |
| Knee | MST | 47/50 | 3/50 |

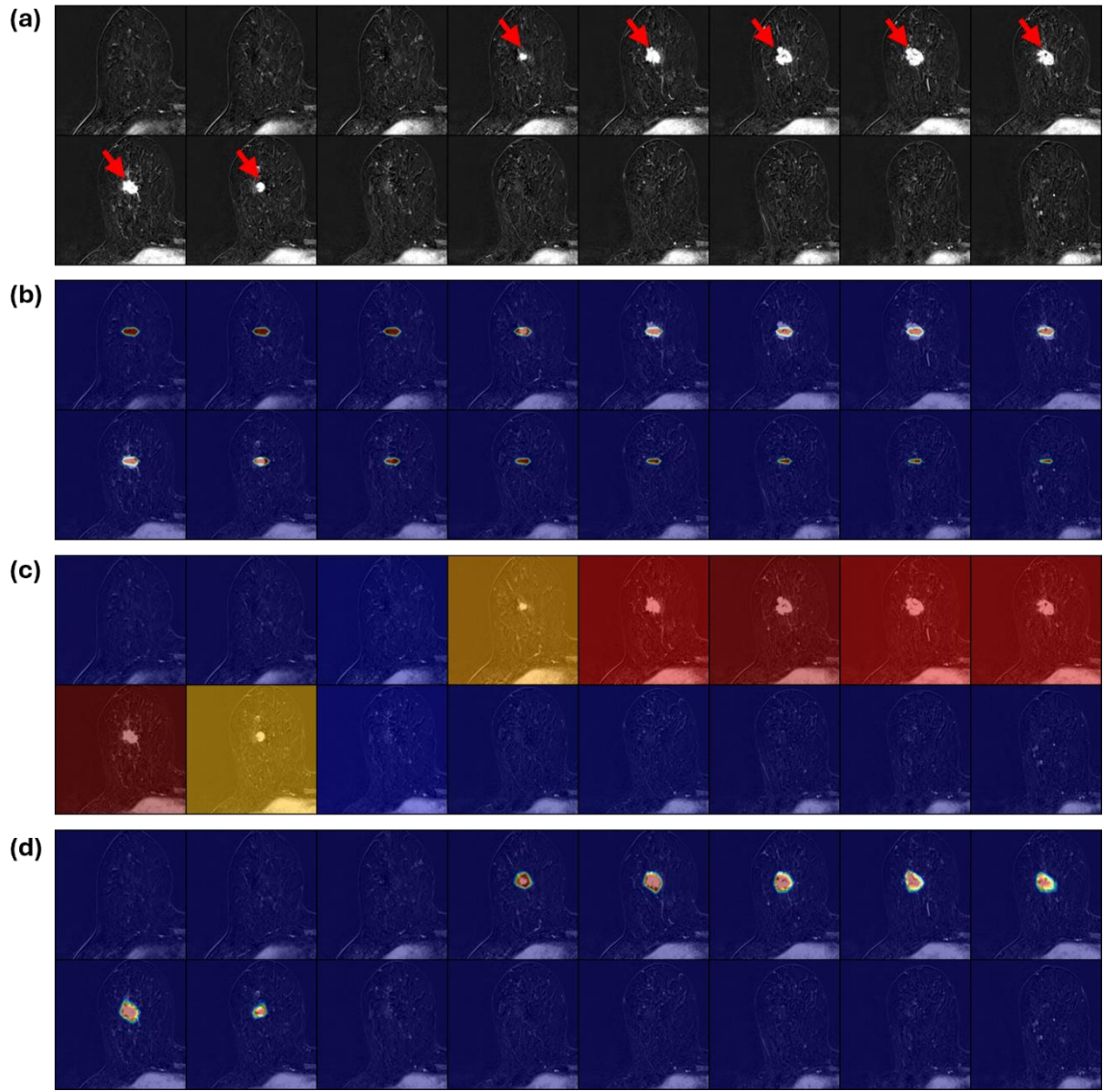

**Figure 4: Exemplary Saliency Maps as a Function of Model Architecture on the Breast MRI Dataset.**
(a) Consecutive axial slices of an MRI scan of the right breast showing subtraction images with a malignant lesion highlighted by red arrows. (b) The saliency maps of the conventional 3D ResNet are spread across the provided image slices and highlight attention on slices without the lesion. The color coding toward blue indicates low attention, while the spectrum toward red indicates high attention. (c) In contrast, the MST Slice Transformer focused exclusively on slices with the lesion. (d) The combined attention map results from two attention mechanisms, i.e., the Slice Transformer's attention to specific slices and the within-slice attention to specific patches. Thereby, precise localization that corresponds well to the lesion is derived.

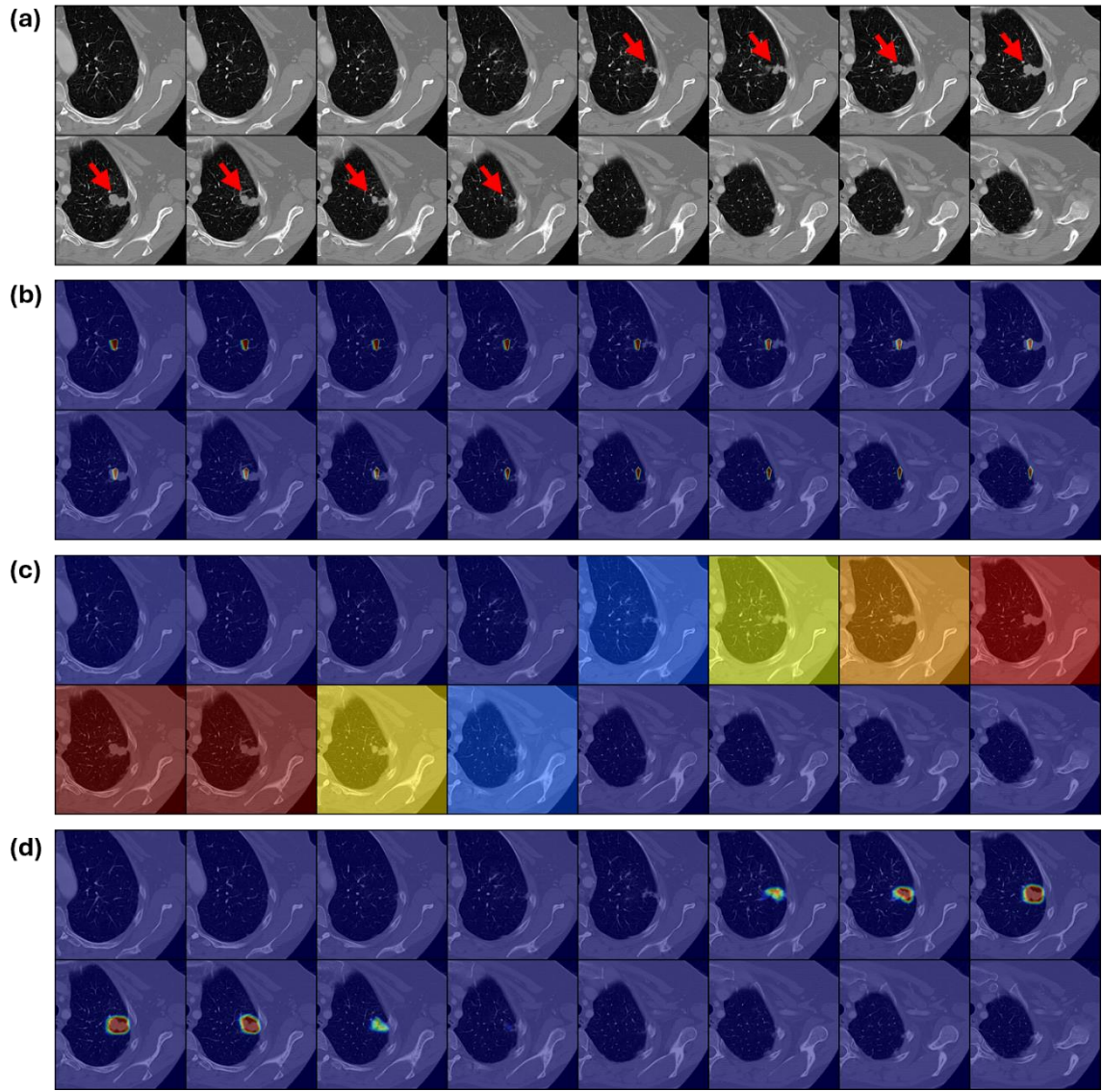

**Figure 5: Exemplary Saliency Maps as a Function of Model Architecture on the Lung CT Dataset.** (a) Consecutive axial slices of the left lung centered around a malignant pulmonary nodule are highlighted by red arrows. (b-d) Image organization as in **Figure 4**.

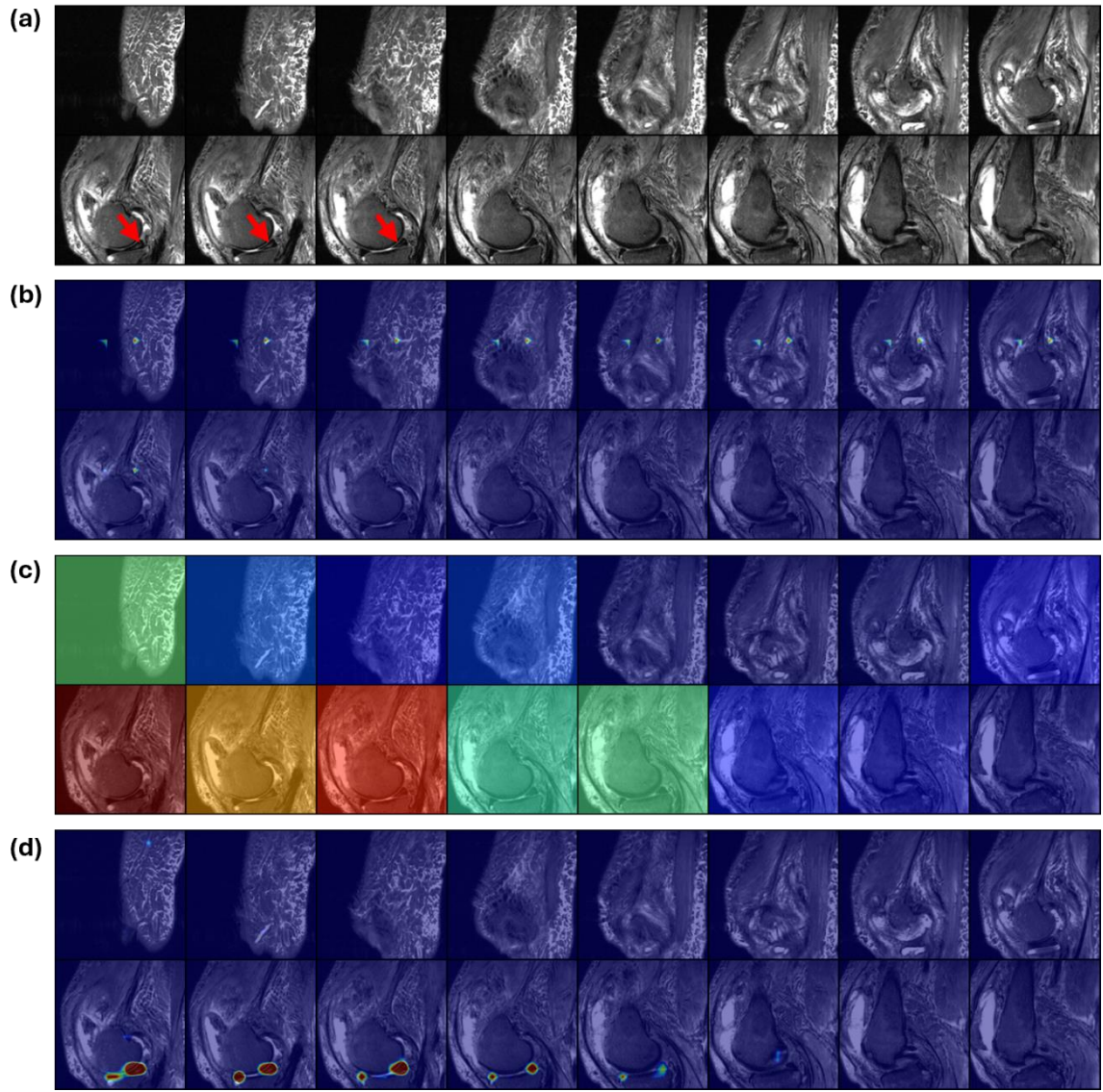

**Figure 6: Exemplary Saliency Maps as a Function of Model Architecture on the Knee MRI Dataset.**
(a) Consecutive sagittal slices of an MRI study (T2-weighted with fat saturation) of a knee with a medial meniscus tear (posterior horn) highlighted by red arrows. (b-d) Image organization as in **Figure 4**.

## Ablation Experiments

We conducted ablation studies to analyze the optimal architecture for the Medical Slice Transformer (MST) framework. First, we evaluated the importance of the Transformer for aggregating slices on the diagnostic performance. Second, we explored different architectures and associated parameters for the encoder.

**Impact of Slice Transformer**

To assess the impact of the Transformer in the MST framework, we analyzed the performance when replacing the Transformer with a linear layer to aggregate the information of all slices. Additionally, we tested the performance when averaging the feature vectors. To assess the impact of positional encoding, we compared the performance of MST models with a variant that uses additive positional encoding. Specifically, each slice was assigned a positional embedding based on its slice number, enabling the model to capture spatial dependencies. For all tests, we used DINOv2 as the backbone.

The Transformer model without positional embedding achieved the highest performance across all three datasets (**Table 2**). Replacing the Transformer with a linear layer or averaging the features resulted in consistently lower AUC values across all datasets.

**Table 2: Classification Performance as a Function of Slice Aggregation and Dataset.** Results are shown as mean accuracy ± standard deviation. The best-performing model for each dataset is highlighted in bold. P-values are given for paired comparisons to the Transformer. Abbreviations: AdPE=Additive Positional Embedding

|  | Breast MRI | Chest CT | Knee MRI |
|---|---|---|---|
| Transformer | **0.94±0.01** | **0.95±0.01** | **0.85±0.04** |
| Transformer (with AdPE) | 0.94±0.01 (P=0.34) | 0.93±0.02 (P=0.17) | 0.83±0.04 (P=0.41) |
| Linear Layer | 0.88±0.02 (P<0.001) | 0.93±0.02 (P=0.26) | 0.78±0.04 (P=0.07) |
| Average | 0.91±0.02 (P=0.03) | 0.92±0.02 (P=0.07) | 0.82±0.04 (P=0.31) |

**Impact of Backbone Architecture**

To further examine the influence of the feature extractor on MST's performance, we conducted experiments with different backbone architectures. We evaluated the effect of increasing the model size using a larger DINOv2 variant ("base") with 86 million parameters instead of 21 million. We also tested augmenting DINOv2 with register tokens (29) to study if this architectural modification would enhance feature representation and overall performance. Additionally, we assessed the impact of freezing the weights of the DINOv2 backbone during training to determine the necessity of fine-tuning. Lastly, we replaced DINOv2 with a pretrained 2D ResNet to compare self-supervised learning features with those obtained from supervised ImageNet pretraining.
The results are summarized in **Table 3**. The larger DINOv2 variant performed comparable to the smaller DINOv2. Incorporating registers into DINOv2 did not yield improvements. Freezing DINOv2 weights led to a significant decrease in AUC values on breast MRI and chest CT (reduced to 0.62 ± 0.03 and 0.66 ± 0.03, respectively), highlighting the necessity of fine-tuning. Replacing DINOv2 with a pretrained ResNet decreased AUC values across all datasets, with a significant drop on the chest CT dataset (0.92 ± 0.02 vs. 0.95 ± 0.01; P = 0.01).

**Table 3: Classification Performance as a Function of the Transformer Model's Backbone Variant and Dataset.** Results are shown as mean accuracy ± standard deviation. The best-performing model for each dataset is highlighted in bold.

|                         | Breast MRI           | Chest CT             | Knee MRI             |
| ----------------------- | -------------------- | -------------------- | -------------------- |
| **MST (DINOv2)**        | **0.94±0.01**        | **0.95±0.01**        | **0.85±0.04**        |
| MST (DINOv2 Base)       | 0.93±0.02 (P=0.69)   | 0.93±0.02 (P=0.27)   | 0.83±0.04 (P=0.41)   |
| MST (DINOv2 Registers)  | 0.93±0.02 (P=0.41)   | 0.92±0.02 (P=0.02)   | 0.82±0.04 (P=0.31)   |
| MST (DINOv2 Frozen)     | 0.62±0.03 (P<0.001)  | 0.66±0.03 (P<0.001)  | 0.79±0.04 (P=0.14)   |
| MST (ResNet)            | 0.93±0.01 (P=0.55)   | 0.92±0.02 (P=0.01)   | 0.78±0.04 (P=0.07)   |

The qualitative evaluation showed that the saliency maps of the MST-DINOv2 model were more focused than ResNet or DINOv2-registers as backbones (see Supplemental **Fig. S1-S4)**.

## Discussion

This study addresses two pivotal challenges in applying deep learning to medical imaging: enhancing diagnostic accuracy and improving model explainability through more accurate saliency maps. By introducing the MST framework, which adapts 2D self-supervised models like DINOv2 to 3D medical images, we demonstrate significant advancements in both areas.

Our MST model achieved superior diagnostic performance across diverse datasets—including breast MRI, chest CT, and knee MRI—evidenced by higher AUC values compared to conventional 3D ResNet architectures. This finding underscores the efficacy of leveraging 2D self-supervised models for 3D medical imaging tasks, extending the benefits observed in previous studies focused on 2D modalities (30–32). Previous studies have demonstrated the effectiveness of slice-wise feature extraction using Transformer architectures for 3D medical image analysis, primarily in brain MRI. However, these efforts did not explore using DINOv2 as a self-supervised feature extraction backbone. Our study demonstrates the effectiveness of DINOv2 across a range of modalities, including breast MRI, knee MRI, and chest CT.

At least of equal importance is the substantial improvement in model explainability. The attention mechanisms inherent in the Transformer architecture provided more precise and informative saliency maps than traditional Grad-CAM methods used with convolutional neural networks. This enhanced localization of pathologies aligns with findings from prior research indicating that Transformer-based methods yield more accurate visual explanations in medical imaging (15,33,34). This improvement is important since improved saliency maps aid in understanding the model's decision-making process and increase the trustworthiness of deep learning models in clinical settings (5–7).

However, limitations need to be acknowledged. The computational demands of processing high-resolution 3D imaging data pose practical challenges: CT scans can have hundreds of slices and spatial resolutions of 512x512 pixels or higher, exceeding current GPU memory capacities. This constraint necessitates downsampling, which may result in the loss of critical diagnostic information. The current study focused on single MRI sequences (i.e., subtraction images and T2-weighted fat-saturated sequences) and unenhanced CT images for this proof-of-concept in medical imaging analysis. In the clinic, multiple contrast phases and reconstructions are used for CT evaluations, and multiple sequences and orientations are used for MRI. Current technical limitations that narrow the input data stream (i.e., the number of input image datasets) may limit the model's ability to fully capture the complex anatomic and pathologic information. Future research is needed to explore the integration of different imaging modalities to enhance model performance and robustness.

In conclusion, adapting 2D self-supervised models like DINOv2 to 3D medical imaging enhances diagnostic accuracy and significantly improves model explainability through better saliency maps. These advancements may provide a cornerstone for overcoming key barriers in deploying deep learning models in clinical practice. Future research should focus on refining these methods and integrating multi-sequence and multi-phase cross-sectional imaging data as well as multimodal data to enhance diagnostic capabilities further.

## Methods

### Dataset Collection and Preprocessing

We used three publicly available datasets to train and test our approach (**Table 4**).

**Table 4**: **Included Datasets and Patient Cohorts.** Data included breast MRI studies from the Duke-Breast-Cancer-MRI (DUKE) dataset, chest CT studies from the Lung Image Database Consortium and Image Database Resource Initiative (LIDC-IDRI) dataset, and knee MRI studies from the MRNet dataset.

| Dataset Name (Reference) | Modality | Anatomic Region | Patient Number [n] (Pre-Exclusion) | Patient Number [n] (Post-Exclusion) |
|---|---|---|---|---|
| DUKE (25) | MRI | Breast | 922 | 651 |
| LIDC-IDRI (26) | CT | Chest | 1,010 | 722 |
| MRNet (27) | MRI | Knee | 1,199 | 1,199 |

The **DUKE dataset** comprises patients with unilateral or bilateral breast cancer acquired at Duke University Medical Center, USA, between January 2000 and March 2014 (25,35). The following sequences and orientations were available for each MRI study: a non-fat saturated, axial T1-weighted sequence and an axial T1-weighted fat-saturated gradient echo sequence (pre-contrast), and typically four post-contrast, axial T1-weighted fat-saturated sequences. Of these, we used the axial T1-weighted pre-contrast and first post-contrast sequences. Each MRI study was separated into left and right breast image stacks, with each side labeled as either cancerous or non-cancerous (i.e., containing a malignant breast lesion or not). We excluded 271 MRI studies where the specific cancer laterality was not provided. Consequently, 651 MRI studies of 651 patients (mean age, 53±12 years [± SD]) were included. These studies were resampled to a resolution of 0.7×0.7×3.0 mm³ to standardize voxel dimensions. Subtraction images were generated by subtracting the first post-contrast images from the pre-contrast images to enhance contrast differences indicative of malignancy versus benign tissue changes. Finally, the images were cropped to a size of 224×224×32 voxels to fit our model input requirements.

The **LIDC-IDRI dataset** contains chest CT scans from 1010 patients with 7371 lung nodules (26). The dataset was acquired by seven academic centers in the United States and eight medical imaging companies using various CT scanners to facilitate the development, training, and evaluation of computer-aided diagnosis systems for lung cancer detection. The CT scans were acquired with slice thicknesses of 0.6 to 5.0 mm (mean 1.9 mm), in-plane pixel sizes of 0.46 to 0.98 mm (mean 0.67 mm), and a resolution of 512x512 voxels with 65 to 764 slices (mean 240). For 928 lung nodules (of 7371), segmentation outlines were provided by all four expert radiologists. We used the pylidc library (36) to aggregate the segmentation outlines by combining annotations based on consensus among the radiologists. More specifically, ground-truth segmentations were established on a per-voxel basis based on a consensus of at least two (of four) expert radiologists (i.e., ≥50% consensus). Practically, the radiologists had delineated the lung nodules and rated the likelihood of malignancy on a 5-point scale ranging from "1: Highly Unlikely" to "5: Highly Suspicious." We excluded 135 patients with nodules smaller than 3 mm because these small nodules are clinically insignificant, challenging to accurately delineate, and may introduce noise. Additionally, 18 nodules were excluded due to their proximity, making them unresolvable by pylidc. The average malignancy rating was calculated for the remaining annotations and on a per-nodule basis. One thousand twenty-six nodules with an average rating of three were excluded because these intermediate ratings represent uncertain cases, which could introduce ambiguity. Nodules with an average rating of four and five were classified as malignant, while those with an average rating of one or two were classified as benign. The chest CT scans were then cropped to a spatial size of 224×224×32 voxels centered around the nodule segmentation outline. Altogether, chest CT images containing 1,625 lung nodules (1,123 benign, 502 malignant) from 722 patients, with an average of two lung nodules/patient, were included.

The **MRNet dataset** was curated by Stanford University to further the development of machine learning for knee MRI (27). The publicly available dataset (without the private test set) comprises MRI studies from 1,199 patients collected between January 2001 and December 2012 using 1.5 T and 3T MRI scanners. It includes 1,250 knee MRI scans with coronal T1-weighted, sagittal T2-weighted fat-saturated, and axial proton density-weighted fat-saturated sequences. Of these, 449 MRI scans (36 %) had meniscal tears. The sagittal T2-weighted fat-saturated sequences were chosen for primary analysis, as they are considered most relevant for evaluating meniscal pathology. The majority vote of three expert MSK radiologists with 6–19 years (average 12 years) of clinical experience was used as the reference label to indicate whether the medial or lateral meniscus was intact or torn. Intact was defined as normal, degenerated without a tear, or post-surgically altered without a tear. Torn was defined as increased signal intensity reaching the articular surface on at least two adjacent slices or a morphologically altered tissue configuration. To standardize the input data for our deep learning model, the images were cropped at a consistent size of 150×150×32 pixels and resampled to 224×224×32 voxels.

## Model Implementation and Training

For the breast MRI and chest CT datasets, we divided the samples into training-validation and test sets with an 80% to 20% split, stratified by label distribution. For the knee MRI dataset, the publicly available validation set (consisting of 120 MRI studies) was used as the test set. The training-validation set was subdivided into 80% training and 20% validation.

We employed ResNet50 (28) with three-dimensional convolutional kernels as our reference model. The model architecture included an input convolutional block, max pooling, four down-sampling convolutional blocks, average pooling, and a final linear layer. The model was implemented using the MONAI (37) library.

Our proposed Medical Slice Transformer (MST) architecture comprised an encoder using the Transformer (11) architecture and an image encoder (**Figure 1**). The Transformer encoder consisted of one layer, with 12 or 16 attention heads, and a feed-forward network with 384, 512, or 768 channels, depending on the image encoder used.

The image encoder was exchangeable in that any model that inputs 2D images and outputs 1D feature vectors can be employed for this task. This study focused on ResNet50, pretrained on ImageNet, and DINOV2 (17). For MST-ResNet, we removed the final linear layer and used the 512-dimensional feature vector obtained after the pooling operation. For DINOV2, we used the 384- or 768- dimensional feature vector of the classification token from the final layer.

The feature vectors from all slices were concatenated with a classification token and fed into the slice Transformer. The output vector of the classification token was then passed through a linear layer to obtain the class logits.

All models were trained using a batch size of 2 and 16-bit precision. We employed weighted sampling to address class imbalances. Furthermore, we used the TorchIO library (38) to apply the following data augmentation techniques to address overfitting: random flip, Gaussian noise, random rotation, and random signal inversion.

The training was terminated if the validation AUC values did not improve over 50 epochs, and the checkpoint with the highest validation AUC value was used for testing. The AdamW(39) optimizer was used with a weight decay of $1e^{-2}$ and learning rates of $1e^{-4}$ for 3D ResNet, $1e^{-5}$ for MST-ResNet, and $1e^{-6}$ for MST-DINOV2. Cross-entropy loss was used as the loss function.

**Saliency Maps**

We employed two techniques to visualize the regions of interest within the images, i.e., those parts of the images the model focused on when making a decision: Gradient-weighted Class Activation Mapping(8) (GradCAM) for convolutional-based models and scaled-dot product attention maps for transformer-based models (12).

*GradCAM*

For the ResNet, we computed gradients for the rectified linear unit function in the last convolutional layer of the ResNet model. The activation outputs were then multiplied by the corresponding gradients to generate saliency maps.

*Attention Map*

For the Transformer-based models, we extracted the scaled-dot product attention from the query-key mechanism in the last layer. We focused on the attention weights of the classification token relative to all other tokens, normalizing the weights so that their sum was one. We then averaged the attention weights across all heads.

For the MST model, we multiplied the attention weights from the Transformer with those provided by the image encoder. We applied linear interpolation to align the attention maps with the input image dimensions.

**Statistical Analysis**

All statistical analyses were performed using Python v3.10 and conducted by G.M.F. and D.T..

The AUC values were calculated using the scikit-learn module (40). To evaluate whether AUC values between models were statistically significant, Delong's test (41) was applied using a fast implementation (42). Confidence intervals for the ROC curves were estimated using the bootstrap method with 1,000 iterations.

P-values < 0.05 were considered statistically significant for all tests.

## Data availability

All datasets in this study are publicly available.

DUKE dataset: https://doi.org/10.7937/TCIA.e3sv-re93;

LIDC-IDRI dataset: https://doi.org/10.7937/K9/TCIA.2015.LO9QL9SX;

MRNet dataset: https://stanfordmlgroup.github.io/competitions/mrnet/.

## Code availability

The code is publicly available at https://github.com/mueller-franzes/MST

## References

1. Lång K, Josefsson V, Larsson AM, Larsson S, Högberg C, Sartor H, et al. Artificial intelligence-supported screen reading versus standard double reading in the Mammography Screening with Artificial Intelligence trial (MASAI): a clinical safety analysis of a randomised, controlled, non-inferiority, single-blinded, screening accuracy study. The Lancet Oncology. 2023 Aug;24(8):936–44.

2. Kuo RYL, Harrison C, Curran TA, Jones B, Freethy A, Cussons D, et al. Artificial Intelligence in Fracture Detection: A Systematic Review and Meta-Analysis. Radiology. 2022 Jul;304(1):50–62.

3. Ching T, Himmelstein DS, Beaulieu-Jones BK, Kalinin AA, Do BT, Way GP, et al. Opportunities and obstacles for deep learning in biology and medicine. J R Soc Interface. 2018 Apr;15(141):20170387.

4. Altaf F, Islam SMS, Akhtar N, Janjua NK. Going Deep in Medical Image Analysis: Concepts, Methods, Challenges, and Future Directions. IEEE Access. 2019;7:99540–72.

5. Singh A, Sengupta S, Lakshminarayanan V. Explainable Deep Learning Models in Medical Image Analysis. J Imaging. 2020 Jun 20;6(6):52.

6. Gulum MA, Trombley CM, Kantardzic M. A Review of Explainable Deep Learning Cancer Detection Models in Medical Imaging. Applied Sciences. 2021 May 17;11(10):4573.

7. Dhar T, Dey N, Borra S, Sherratt RS. Challenges of Deep Learning in Medical Image Analysis—Improving Explainability and Trust. IEEE Trans Technol Soc. 2023 Mar;4(1):68–75.

8. Selvaraju RR, Cogswell M, Das A, Vedantam R, Parikh D, Batra D. Grad-CAM: Visual Explanations from Deep Networks via Gradient-Based Localization. In: 2017 IEEE International Conference on Computer Vision (ICCV) [Internet]. Venice: IEEE; 2017 [cited 2024 Aug 17]. p. 618–26. Available from: http://ieeexplore.ieee.org/document/8237336/

9. Shi Y, Wang C, Liu D, Cai W, Cabezas M. Understanding the Role of Saliency Maps for Biomarker Research in 3D Medical Imaging Classification. In: 2023 International Conference on Digital Image

Computing: Techniques and Applications (DICTA) [Internet]. Port Macquarie, Australia: IEEE; 2023 [cited 2024 Aug 29]. p. 41–8. Available from: https://ieeexplore.ieee.org/document/10410916/

10. Wiśniewski M, Giulivi L, Boracchi G. SE3D: A Framework For Saliency Method Evaluation In 3D Imaging [Internet]. arXiv; 2024 [cited 2024 Aug 29]. Available from: https://arxiv.org/abs/2405.14584

11. Vaswani A, Shazeer N, Parmar N, Uszkoreit J, Jones L, Gomez AN, et al. Attention Is All You Need. arXiv:170603762 [cs] [Internet]. 2017 Dec 5 [cited 2021 Mar 29]; Available from: http://arxiv.org/abs/1706.03762

12. Dosovitskiy A, Beyer L, Kolesnikov A, Weissenborn D, Zhai X, Unterthiner T, et al. An Image is Worth 16x16 Words: Transformers for Image Recognition at Scale. arXiv:201011929 [cs] [Internet]. 2021 Jun 3 [cited 2022 Mar 28]; Available from: http://arxiv.org/abs/2010.11929

13. Nishigaki D, Suzuki Y, Watabe T, Katayama D, Kato H, Wataya T, et al. Vision transformer to differentiate between benign and malignant slices in 18F-FDG PET/CT. Sci Rep. 2024 Apr 9;14(1):8334.

14. Park S, Kim G, Oh Y, Seo JB, Lee SM, Kim JH, et al. Self-evolving vision transformer for chest X-ray diagnosis through knowledge distillation. Nat Commun. 2022 Jul 4;13(1):3848.

15. Wollek A, Graf R, Čečatka S, Fink N, Willem T, Sabel BO, et al. Attention-based Saliency Maps Improve Interpretability of Pneumothorax Classification. Radiology: Artificial Intelligence. 2023 Mar 1;5(2):e220187.

16. Ouali Y, Hudelot C, Tami M. An Overview of Deep Semi-Supervised Learning [Internet]. arXiv; 2020 [cited 2024 Aug 29]. Available from: https://arxiv.org/abs/2006.05278

17. Oquab M, Darcet T, Moutakanni T, Vo H, Szafraniec M, Khalidov V, et al. DINOv2: Learning Robust Visual Features without Supervision [Internet]. arXiv; 2023 [cited 2024 Aug 16]. Available from: https://arxiv.org/abs/2304.07193

18. Huix JP, Ganeshan AR, Haslum JF, Söderberg M, Matsoukas C, Smith K. Are Natural Domain Foundation Models Useful for Medical Image Classification? In: 2024 IEEE/CVF Winter Conference on Applications of Computer Vision (WACV) [Internet]. Waikoloa, HI, USA: IEEE; 2024 [cited 2024 Aug 27]. p. 7619–28. Available from: https://ieeexplore.ieee.org/document/10483777/

19. Huang Y, Zou J, Meng L, Yue X, Zhao Q, Li J, et al. Comparative Analysis of ImageNet Pre-Trained Deep Learning Models and DINOv2 in Medical Imaging Classification [Internet]. arXiv; 2024 [cited 2024 Aug 27]. Available from: https://arxiv.org/abs/2402.07595

20. Paderno A, Rau A, Bedi N, Bossi P, Mercante G, Piazza C, et al. Computer Vision Foundation Models in Endoscopy: Proof of Concept in Oropharyngeal Cancer. The Laryngoscope. 2024 Jun 8;lary.31534.

21. Shakouri M, Iranmanesh F, Eftekhari M. DINO-CXR: A self supervised method based on vision transformer for chest X-ray classification [Internet]. arXiv; 2023 [cited 2024 Jul 1]. Available from: https://arxiv.org/abs/2308.00475

22. Tayebi Arasteh S, Misera L, Kather JN, Truhn D, Nebelung S. Enhancing diagnostic deep learning via self-supervised pretraining on large-scale, unlabeled non-medical images. Eur Radiol Exp. 2024 Feb 8;8(1):10.

23. Truong T, Mohammadi S, Lenga M. How Transferable Are Self-supervised Features in Medical Image Classification Tasks? 2021 [cited 2024 Jul 1]; Available from: https://arxiv.org/abs/2108.10048

24. Nielsen M, Wenderoth L, Sentker T, Werner R. Self-Supervision for Medical Image Classification: State-of-the-Art Performance with ~100 Labeled Training Samples per Class. Bioengineering. 2023 Jul 28;10(8):895.

25. Saha A, Harowicz MR, Grimm LJ, Weng J, Cain EH, Kim CE, et al. Dynamic contrast-enhanced magnetic resonance images of breast cancer patients with tumor locations [Internet]. The Cancer Imaging Archive; 2022 [cited 2024 Oct 6]. Available from: https://www.cancerimagingarchive.net/collection/duke-breast-cancer-mri/

26. Armato III SG, McLennan G, Bidaut L, McNitt-Gray MF, Meyer CR, Reeves AP, et al. Data From LIDC-IDRI [Internet]. The Cancer Imaging Archive; 2015 [cited 2024 Oct 6]. Available from: https://www.cancerimagingarchive.net/collection/lidc-idri/

27. Bien N, Rajpurkar P, Ball RL, Irvin J, Park A, Jones E, et al. Deep-learning-assisted diagnosis for knee magnetic resonance imaging: Development and retrospective validation of MRNet. Saria S, editor. PLoS Med. 2018 Nov 27;15(11):e1002699.

28. He K, Zhang X, Ren S, Sun J. Deep Residual Learning for Image Recognition. In: 2016 IEEE Conference on Computer Vision and Pattern Recognition (CVPR) [Internet]. Las Vegas, NV, USA: IEEE; 2016 [cited 2023 Jun 4]. p. 770–8. Available from: http://ieeexplore.ieee.org/document/7780459/

29. Darcet T, Oquab M, Mairal J, Bojanowski P. Vision Transformers Need Registers [Internet]. arXiv; 2023 [cited 2024 Aug 16]. Available from: https://arxiv.org/abs/2309.16588

30. Jun E, Jeong S, Heo DW, Suk HI. Medical Transformer: Universal Encoder for 3-D Brain MRI Analysis. IEEE Trans Neural Netw Learning Syst. 2023;1–11.

31. Alp S, Akan T, Bhuiyan MdS, Disbrow EA, Conrad SA, Vanchiere JA, et al. Joint transformer architecture in brain 3D MRI classification: its application in Alzheimer's disease classification. Sci Rep. 2024 Apr 18;14(1):8996.

32. Jang J, Hwang D. M3T: three-dimensional Medical image classifier using Multi-plane and Multi-slice Transformer. In: 2022 IEEE/CVF Conference on Computer Vision and Pattern Recognition (CVPR) [Internet]. New Orleans, LA, USA: IEEE; 2022 [cited 2024 Sep 21]. p. 20686–97. Available from: https://ieeexplore.ieee.org/document/9878673/

33. Murphy ZR, Venkatesh K, Sulam J, Yi PH. Visual Transformers and Convolutional Neural Networks for Disease Classification on Radiographs: A Comparison of Performance, Sample Efficiency, and Hidden Stratification. Radiology: Artificial Intelligence. 2022 Nov 1;4(6):e220012.

34. Komorowski P, Baniecki H, Biecek P. Towards Evaluating Explanations of Vision Transformers for Medical Imaging. In: 2023 IEEE/CVF Conference on Computer Vision and Pattern Recognition Workshops (CVPRW) [Internet]. Vancouver, BC, Canada: IEEE; 2023 [cited 2024 Sep 26]. p. 3726–32. Available from: https://ieeexplore.ieee.org/document/10208407/

35. Saha A, Harowicz MR, Grimm LJ, Kim CE, Ghate SV, Walsh R, et al. A machine learning approach to radiogenomics of breast cancer: a study of 922 subjects and 529 DCE-MRI features. Br J Cancer. 2018 Aug;119(4):508–16.

36. Hancock MC, Magnan JF. Lung nodule malignancy classification using only radiologist-quantified image features as inputs to statistical learning algorithms: probing the Lung Image Database Consortium dataset with two statistical learning methods. J Med Imag. 2016 Dec 8;3(4):044504.

37. Cardoso MJ, Li W, Brown R, Ma N, Kerfoot E, Wang Y, et al. MONAI: An open-source framework for deep learning in healthcare. 2022 [cited 2023 Jul 13]; Available from: https://arxiv.org/abs/2211.02701

38. Pérez-García F, Sparks R, Ourselin S. TorchIO: A Python library for efficient loading, preprocessing, augmentation and patch-based sampling of medical images in deep learning. 2020 [cited 2023 Apr 13]; Available from: https://arxiv.org/abs/2003.04696

39. Loshchilov I, Hutter F. Decoupled Weight Decay Regularization. 2017 [cited 2024 Jan 10]; Available from: https://arxiv.org/abs/1711.05101

40. Pedregosa F, Varoquaux G, Gramfort A, Michel V, Thirion B, Grisel O, et al. Scikit-learn: Machine learning in Python. Journal of Machine Learning Research. 2011;12:2825–30.

41. DeLong ER, DeLong DM, Clarke-Pearson DL. Comparing the Areas under Two or More Correlated Receiver Operating Characteristic Curves: A Nonparametric Approach. Biometrics. 1988 Sep;44(3):837.

42. Sun X, Xu W. Fast Implementation of DeLong's Algorithm for Comparing the Areas Under Correlated Receiver Operating Characteristic Curves. IEEE Signal Process Lett. 2014 Nov;21(11):1389–93.

43. Virtanen P, Gommers R, Oliphant TE, Haberland M, Reddy T, Cournapeau D, et al. SciPy 1.0: fundamental algorithms for scientific computing in Python. Nat Methods. 2020 Mar 2;17(3):261–72.

## Acknowledgements

- The data used in this publication was managed using the research data management platform Coscine with storage space granted by the Research Data Storage (RDS) of the DFG and Ministry of Culture and Science of the State of North Rhine-Westphalia (DFG: INST222/1261-1 and MKW: 214-4.06.05.08 - 139057).
- The authors gratefully acknowledge the computing time provided to them at the NHR Center NHR4CES at RWTH Aachen University (project number p0021834). This is funded by the Federal Ministry of Education and Research, and the state governments participating on the basis of the resolutions of the GWK for national high performance computing at universities (www.nhr-verein.de/unsere-partner).
- The authors acknowledge the National Cancer Institute and the Foundation for the National Institutes of Health, and their critical role in the creation of the free publicly available LIDC/IDRI Database used in this study.

## Funding

- JNK is supported by the German Federal Ministry of Health (DEEP LIVER, ZMVI1-2520DAT111), the Max-Eder-Programme of the German Cancer Aid (grant #70113864), the German Federal Ministry of Education and Research (PEARL, 01KD2104C; CAMINO, 01EO2101; SWAG, 01KD2215A; TRANSFORM LIVER, 031L0312A; TANGERINE, 01KT2302 through ERA-NET Transcan), the German Academic Exchange Service (SECAI, 57616814), the German Federal Joint Committee (Transplant.KI, 01VSF21048) the European Union's Horizon Europe and innovation programme (ODELIA, 101057091; GENIAL, 101096312) and the National Institute for Health and Care Research (NIHR, NIHR213331) Leeds Biomedical Research Centre.
- The views expressed are those of the author(s) and not necessarily those of the NHS, the NIHR or the Department of Health and Social Care.
- DT is supported by the German Federal Ministry of Education and Research (SWAG, 01KD2215A; TRANSFORM LIVER), the European Union's Horizon Europe and innovation programme (ODELIA, 101057091).

## Conflict of Interest

- DT received honoraria for lectures by Bayer, GE, and Philips and holds shares in StratifAI GmbH, Germany and in Synagen GmbH, Germany.
- FK holds shares in StratifAI GmbH
- JNK declares consulting services for Bioptimus, France; Owkin, France; DoMore Diagnostics, Norway; Panakeia, UK; AstraZeneca, UK; Mindpeak, Germany; and MultiplexDx, Slovakia. Furthermore, he holds shares in StratifAI GmbH, Germany, Synagen GmbH, Germany, and has received a research grant by GSK, and has received honoraria by AstraZeneca, Bayer, Daiichi Sankyo, Janssen, Merck, MSD, BMS, Roche, Pfizer and Fresenius.

## Ethics declarations

This study was carried out in accordance with the Declaration of Helsinki.

# Supplementary Material

**Saliency Maps Ablation Experiments**

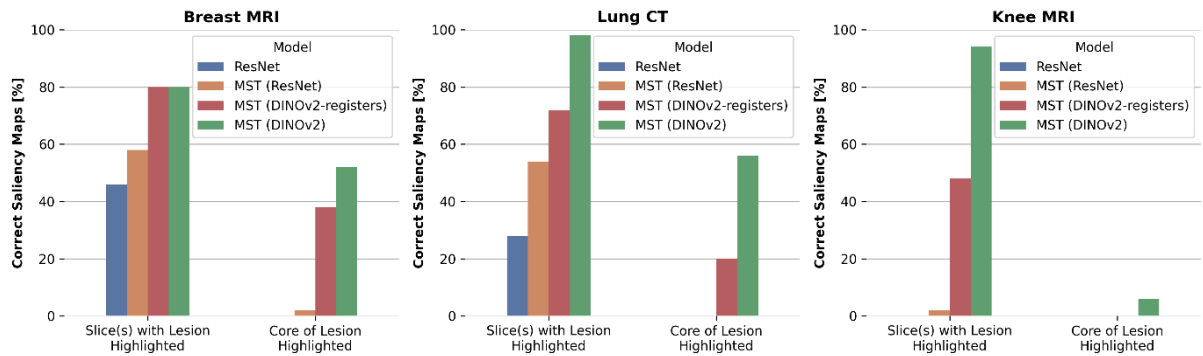

**Figure S1: Quantification of Saliency Map-Lesion-Correspondence as a Function of Imaging Dataset and Model Architecture.** Percentages of (blinded) radiologist evaluation in terms of slice correctness ("Does the saliency map highlight the slice(s) containing the lesion?" - yes/no) and lesion correctness ("Does the saliency map accurately point to the core of the lesion?" - yes/no).

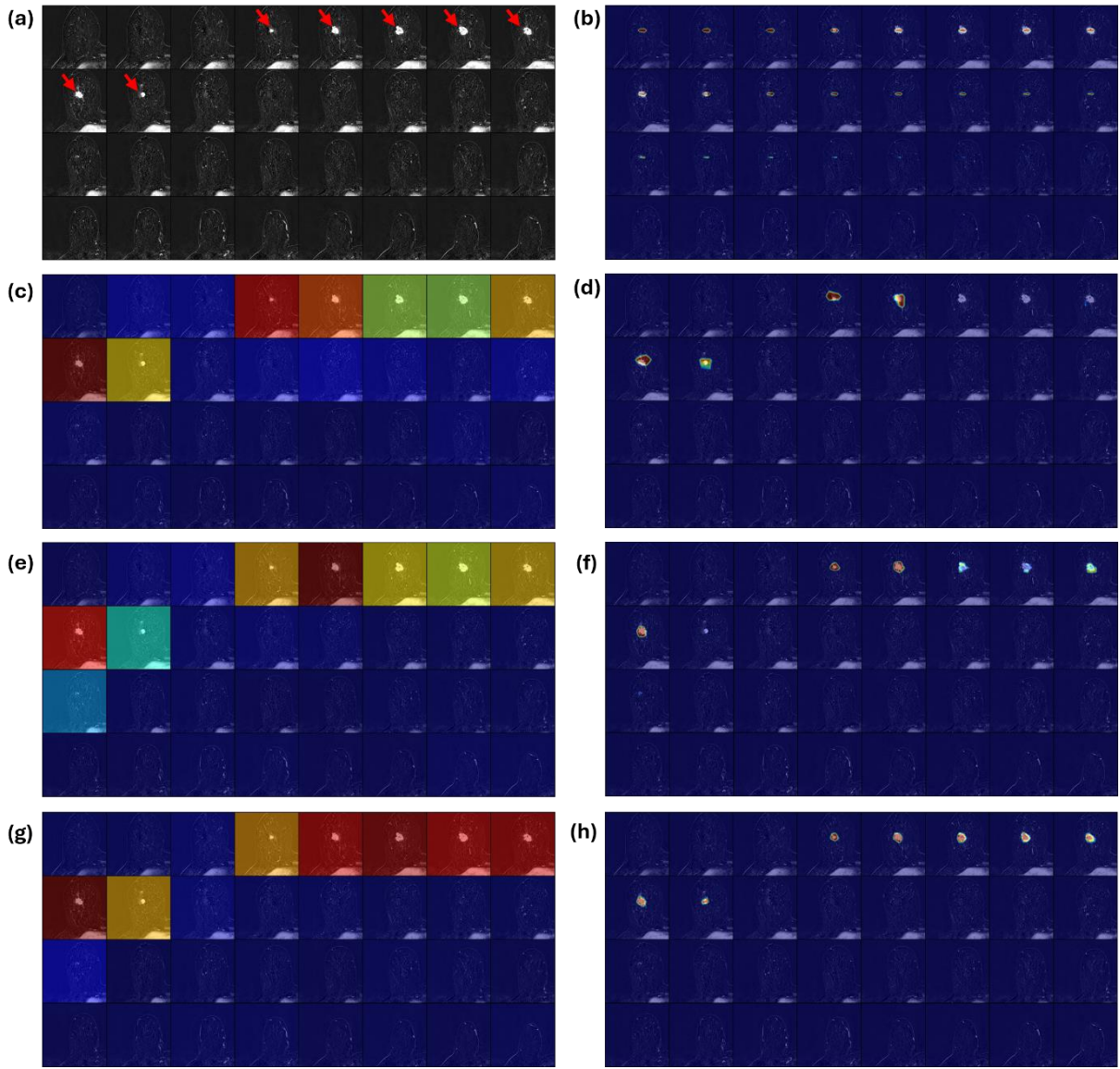

**Figure S2: Saliency Maps as a Function of Model Architecture on the Breast MRI Dataset.** (a) Consecutive axial slices of an MRI scan of the right side of the breast showing subtraction images with a malignant lesion highlighted by red arrows. (b) The saliency maps of the conventional 3D ResNet are spread across all provided sections of the dataset and highlight attention on slices without the lesion. The color coding toward blue indicates low attention, while the spectrum toward red indicates high attention. In contrast, the MST framework focuses on the lesion when using ResNet (c), DINOv2-registers (e), or DINOv2 (g) as an image encoder. The combined attention map results from two attention mechanisms, i.e., the Slice Transformer's attention to specific slices and the within-slice attention to specific patches when using ResNet (d), DINOv2-registers (f) or DINOv2 (h) as image encoder. The precision of lesion localization improves as a function of image encoders, with DINOv2 showing the highest precision.

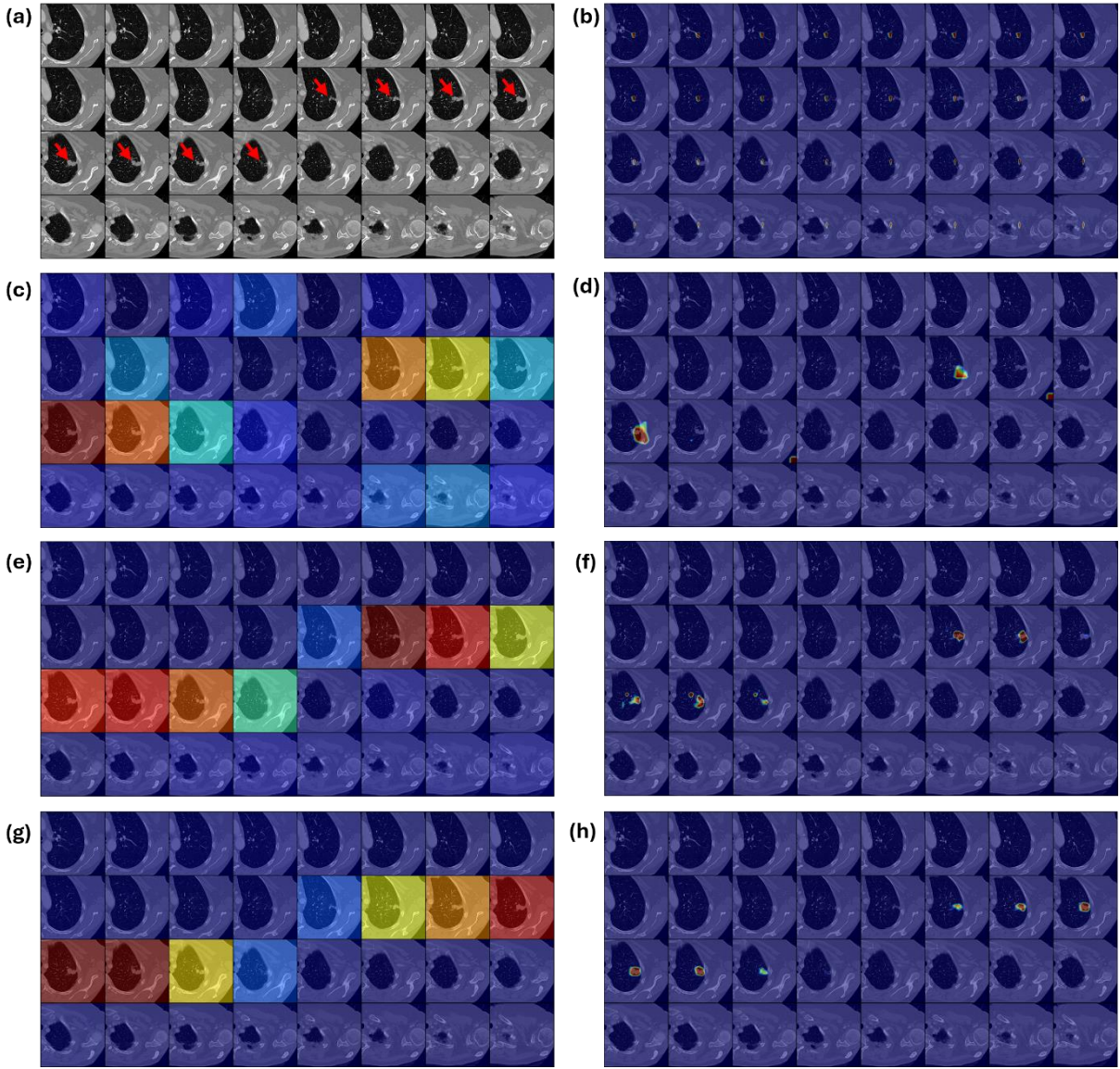

**Figure S3: Saliency Maps as a Function of Model Architecture on the Lung CT Dataset.** (a) Consecutive axial slices of a left sided lung CT scan centered around a large pulmonary nodule highlighted by red arrows. (b-h) Image organization as in **Figure S2**.

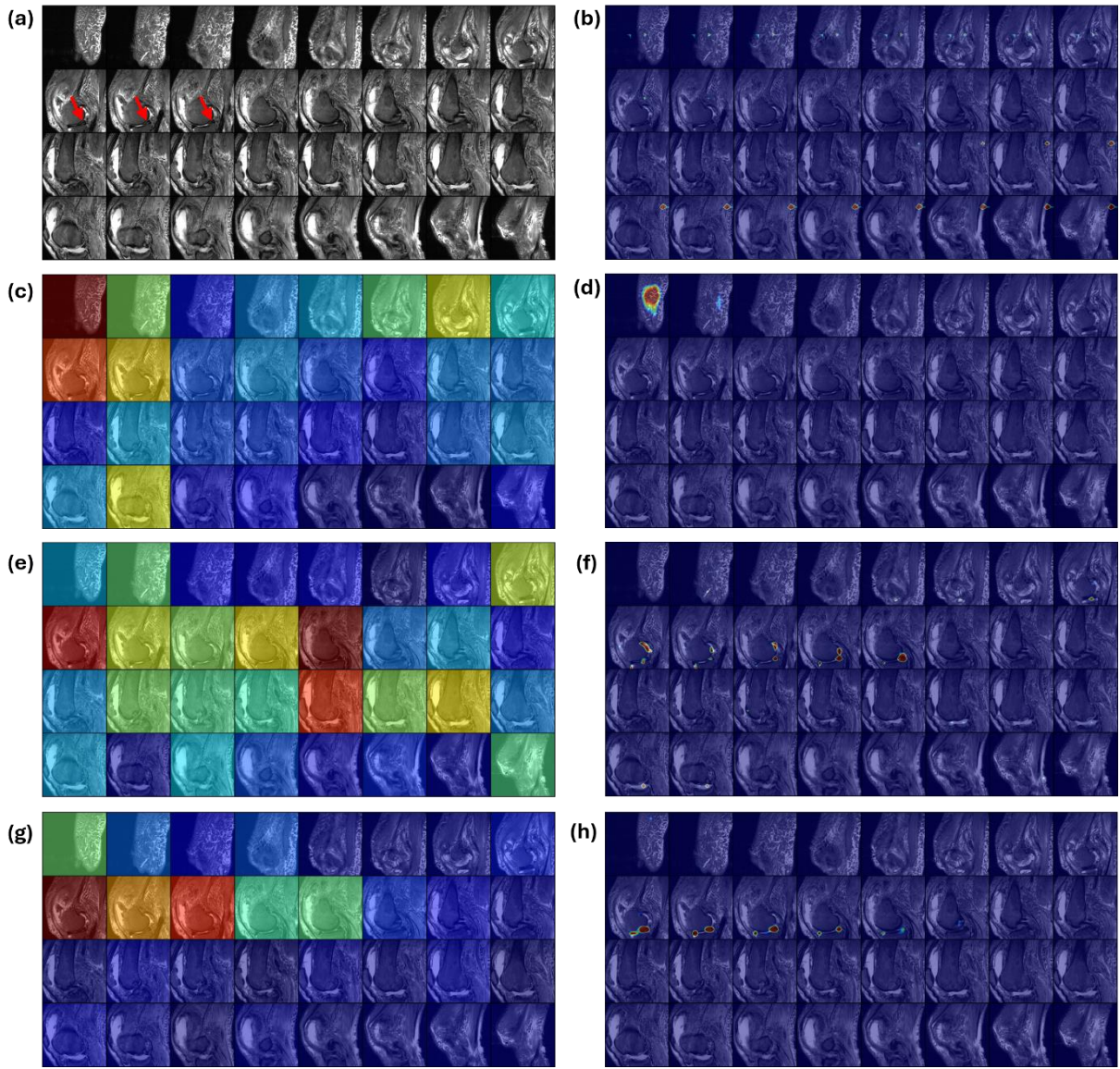

**Figure S4: Saliency Maps as a Function of Model Architecture on the Knee MRI Dataset.** (a) Consecutive sagittal slices of an MRI scan of a knee with meniscus tear highlighted by red arrows. (b-h) Image organization as in **Figure S2**.

**Table S1: Classification Performance as a Function of the ResNet Variant and Dataset.** Results are shown as mean accuracy ± standard deviation. The best-performing model for each dataset is highlighted in bold.

|             | Breast MRI         | Chest CT           | Knee MRI              |
|-------------|--------------------|--------------------|-----------------------|
| **ResNet50**  | **0.91±0.02**    | **0.92±0.02**    | 0.69±0.05             |
| **ResNet18**  | 0.91±0.02 (P=0.63) | 0.91±0.02 (P=0.42) | 0.63±0.05 (P=0.18)    |
| **ResNet101** | 0.90±0.02 (P=0.40) | 0.92±0.02 (P=0.97) | **0.70±0.05** (P=0.72) |